# Atmospheric Response to Fukushima Daiichi NPP (Japan) Accident Reviled by Satellite and Ground observations

Dimitar Ouzounov [1,2], Sergey Pulinets [3,4], Katsumi Hattori[5], Menas Kafatos[1], Patrick Taylor[2]

[1] *Chapman University, One University Drive, Orange, CA 92866, USA*
[2] *NASA Goddard Space Flight Center, Greenbelt, MD 20771, USA*
[3] *Institute of Applied Geophysics, Rostokinskaya str., 9, Moscow, 129128, Russia*
[4] *Space Research Institute RAS, Profsoyuznaya str. 84/32, Moscow 117997, Russia*
[5] *Department of Earth Sciences, Chiba University, Chiba 263-8522, Japan*

Correspondence to: D.Ouzounov (Ouzounov@chapman.edu)

**Abstract**
Immediately after the March 11, 2011 earthquake and tsunami in Japan we started to continuously survey the Outgoing Long-wavelength Radiation (OLR, 10-13 microns) from NOAA/AVHRR. Our preliminary results show the presence of hot spots on the top of the atmosphere over the Fukushima Daiichi Nuclear Power Plant (FDNPP) and due to their persistence over the same region they are most likely not of meteorological origin. On March 14 and 21 we detected a significant increase in radiation (14 W/m$^2$) at the top of the atmosphere which also coincides with a reported radioactivity gas leaks from the FDNPP. After March 21 the intensity of OLR started to decline, which has been confirmed by ground radiometer network. We hypothesize that this increase in OLR was a result of the radioactive leaks released in atmosphere from the FDNPP. This energy triggers ionization of the air near the ground and lead to release of latent heat energy due to change of air humidity and temperature. Our early results demonstrate the potential of the latest development in atmospheric sciences and space-borne observations for monitoring nuclear accidents.

**Introduction**
The March 11th earthquake in Japan triggered extremely destructive tsunami waves of up to 38.9 meters that struck Japan, making it the most expensive natural disaster on record. In response to the disaster Japan Nuclear Energy authorities have announced they turned off 11 nuclear units without causing any damage. However, the tsunami disaster of March 11, 2011 produced significant damage to the Fukushima-Daichi Nuclear Power reactor resulted from a failure of the cooling system. Record show that in
in July 2007, there had been an another incident at the largest nuclear power plant in the world, the Kashiwazaki-Karim this occurred after a strong earthquake near Niigata City. With destruction similar to the Fukushima nuclear power plant, that is a leak of contaminated water and release of radioactive gas into the atmosphere. This single incident of radiation release was known only to a few specialists (Cyranoski, 2007). The tsunami disaster of March 11, 2011 produced significant damage to the FDPP reactors, in particular, the formation of cracks in the reactor containment units 1 and 2 this resulted from a failure of the cooling system. As a result, there have been several explosions of hydrogen, which led to the release into the atmosphere of decay products, such as radioactive cesium and iodine . Given that the adverse consequences of the accident scenario could be global in nature, it was very important that an independent source of information was available to evaluate the level of radioactive contamination. Information about the thermal radiation activities over the FDNPP, could be obtained by using a modification of the technique of registration emitted infrared longwave radiation (OLR), as described in (Ouzounov et al, 2007). The study of ionization of the surface air radon in preparation for the earthquake revealed that the most likely cause for the observed transient thermal radiation anomalies observed before earthquakes are the result of release of latent heat of evaporation of hydration of ions formed by ionization of atmospheric gases radon (Pulinets and Ouzounov, 2011)

**Data Observation and Analysis**

During March 2011 the NOAA/AVHRR (Advanced Very High Resolution Radiometer) satellite recorded OLR radiation over Japan. We noted an anomalous OLR hot spot over the Fukushima Daiichi Nuclear Power Plant (FDNPP) from NOAA/AVHRR satellite data. On March 14 shortly after the March 11 tsunami we detected a strong increase in emitted radiation at the top of the atmosphere (Ouzounov et al., 2011b). We interpreted this as an increase in OLR due to radioactive gas released in the atmosphere from the FDNPP. Similar studies of the physical link between the atmospheric and ionospheric have been previously examined for the historical nuclear accidents at Chernobyl (USSR, 1977) and Three Mile Island (USA, 1979) by Pulinets and Boyarchuk (2004) and Laverov et al. (2011).

OLR is one of the main parameters characterizing the earth's radiation environment. They formed at the top of the atmosphere from an integration of the emissions from the ground, lower atmosphere and clouds (Ohring, G. and Gruber, 1982). OLR are primarily used to study the Earths radiative budget and climate (Gruber, A. and Krueger, 1984; Mehta, A., and J. Susskind, 1999). In this study we analyzed NOAA/AVHRR OLR data during March 2011 over FDNPP. The National Oceanic and Atmospheric Administration (NOAA) Climate Prediction Center (http://www.cdc.noaa.gov/) provide daily and monthly OLR data. However OLR is not directly measured, but is calculated by integrating measurements between 10 and 13 μm using an algorithm (Gruber and Krueger, 1984). These data are mainly sensitive to near surface and cloud temperatures. A daily mean value covering a significant area of the Earth ($90^o$ N- $90^o$ S, $0^o$ E to $357.5^o$ E) with a spatial resolution of $2.5^o$ x $2.5^o$ was used to study the OLR variability in the zone of earthquake activity (Liu, 2000; Ouzounov et al., 2007, Xiong at al, 2010). An increase in radiation and a transient change in OLR were proposed to be related to thermodynamic processes in the atmosphere over seismically active regions (Ouzounov et al, 2007, Pulinets and Ouzounov, 2011).

We started continuously to survey OLR from the NOAA/AVHRR satellite over Japan immediately after the March 11 earthquake and tsunami. Our preliminary results from analysis of several atmospheric parameters over Sendai region show evidence for an existence of pre earthquake signals associated with the March 11[th] M9.0 Great Tohoku earthquake (Ouzounov et al, 2011, a, b and c). To improve the OLR detection capabilities over the region of FDNPP we constructed a new change detection ration known as the temporal coherency (TC_index) in the same manner as that defined for an anomalous thermal field by Tramutoli et al. (1999) and Ouzounov et al. (2007). The TC_index represents the maximum change in the rate of OLR for a specific location over a pre-defined time period. The OLR reference field was computed as the average daily mean from all data for 2011. On March 12, the first transient OLR anomalous field was observed near the FDNPP (Figs 2 and 3). The first strong indication of the formation of a transient OLR anomaly was detected on March 14 (Fig 2) three days after the tsunami damaged the FDNPP and radioactive leaks had been reported March 3-15 (Fig.3). A cumulative graph showing the variation in the OLR TC index is given in Fig. 3. On March 21 OLR reached a maximum value of 14W/m^2 which is 7 times bigger then the normal level of 2 W/m^2 for the same period of 2010. (Fig.3 dash red line). The presence of abnormal OLR in atmosphere coincides by time and space with the ground radioactivity measurements from Stations #31 (Futaba County Namie town Tsushima Nakioki) at 30Km NW from FDNPP (see Fig.1) (MEXT, Reference 2). With values of 110MSv/h at 10.00 LT this maximum also coincides with the air temperature maximum of 8C$^o$ measured near Fukushima city at 07.30LT (Fig.3 blue curve). The Temperature differences between year 2011 and 2010 reviled an temperature anomaly of +6 C$^o$ on March 15$^{th}$ which coincides with the reports of March 13-15 radioactive gas releases reported by TEPCO. On Fig.3 with purple color is marked the time the FDNPP explosions and gas releases in atmosphere reported by TEPCO (Reference 16).

The abnormal air temperature increases in atmosphere matching precisely the time of satellite anomalies and reported almost 50 km NW from FDNPP are ruling out the possibility for alterative explanation of OLR change on the top of the atmosphere as a direct heating processes from the damaged reactor units. Rapid ionization processes could explain the transient increase in the air temperature, triggered by radioactive leaks, which lead to anomalous increase flux of the latent heat over the area of radioactive release into the atmosphere.

This paper demonstrated the possibility of independent survey of radioactive pollution of the environment through the monitoring, at satellite altitude, of thermal anomalies. Despite the low resolution of the AVHRR (2.5 degrees), this methodology allowed reliable registration of thermal anomalies from a relatively small area. This also suggests that using radiometers with a higher resolution would increase the monitoring ability of our method.

**Conclusions**

Our preliminary results show an unusual presence of hot spots on the top of the atmosphere over the radioactive leaks at FDNPP which also been conformed with abnormal change of in situ observation of radiation and atmospheric temperature near FDNPP. We interpreted this abnormal increase in OLR due to radioactive gas being released in atmosphere from the FDNPP. A physical mechanism for generating such anomalies could be the ionization of the atmospheric boundary layer similar to the case of thermal anomalies detected before strong earthquakes (Pulinets and Ouzounov, 2011). The rate of OLR increase over the FDNPP, according to satellite NOAA/AVHRR, reached 14 W/m$^2$, which is comparable with the values recorded before some strong earthquakes (Ouzounov et al, 2007). These findings demonstrate the potential for space-borne observations in monitoring major environmental disasters.

**Acknowledgments**

We wish to thank to Chapman University and NASA Godard Space Flight Center for their kind support. We also thank NOAA/ National Weather Service National Centers for Environmental Prediction Climate Prediction Center for providing OLR data. The Weather Data for Honshu, Japan are provided by Chiba University, Japan. The ground radioactivity level data are obtained from Ministry of Education Culture, Sport Science and Technology (MEXT) Japan. FDNPP information for gas releases in atmosphere data are obtained from TEPCO, Japan

**REFERENCES**
Cyranoski D (2007) Quake shuts world's largest nuclear plant, Nature, 448,26
http://www.mext.go.jp/english/
Gruber, A. and Krueger, A., The status of the NOAA outgoing longwave radiation dataset. *Bulletin of the American Meteorological Society*, 65, 958–962,1984
Laverov N, Pulinets S., Ouzounov D Using thermal ionization effect for remote diagnostics of the radioactive contamination of the environment, Russian Academy of Sciences, 2011 (submitted)
Liu, D.: Anomalies analyses on satellite remote sensing OLR before Jiji earthquake of Taiwan Province, Geo-Information Science,2(1), 33–36, (in Chinese with English abstract),2000
Mehta, A., and J. Susskind, Outgoing Longwave Radiation from the TOVS Pathfinder Path A Data Set, *J. Geophys. Res.,* 104, NO. D10,12193-12212,1999.
Ohring, G. and Gruber, A.: Satellite radiation observations and climate theory, *Advance in Geophysics*., 25, 237–304, 1982.
Ouzounov D., D. Liu, C. Kang , G. Cervone, M. Kafatos, P. Taylor, (2007) Outgoing Long Wave Radiation Variability from IR Satellite Data Prior to Major Earthquakes, *Tectonophysics*, 431, 1-4 , 20, 211-220
Ouzounov D., K.Hattori, S. Pulinets, T. Liu, M. Kafatos, P. Taylor, F. Yang, K. Oyama, S. Kon, (2011b) Integrated Sensing, Analysis and Validation of Atmospheric Signals Associated with Major Earthquakes , *Geophysical Research Abstracts* Vol. 13, EGU2011-11932-1, 2011, EGU General Assembly
Ouzounov D., S. Pulinets, A. Romanov, A. Romanov Jr., K. Tsybulya, D. Davidenko, M. Kafatos and P. Taylor  (2011c) Atmosphere-Ionosphere Response to the M9 Tohoku Earthquake Revealed by Joined Satellite and Ground Observations. Preliminary results; *Geophysical Research Abstracts* Vol. 13, EGU2011, EGU General Assembly, *http://arxiv.org/abs/1105.2841*
Pulinets S, Ouzounov D., 2010, Abstract NH24A-04 presented at 2010 Fall Meeting, AGU, San Francisco, Calif., 13-17 Dec.
Pulinets S. A., Boyarchuk K. A. (2004) Ionospheric Precursors of Earthquakes, Springer, Germany, 315 p.
Pulinets, S. and D. Ouzounov (2011) Lithosphere-Atmosphere-Ionosphere Coupling (LAIC) model - an unified concept for earthquake precursors validation, *Journal of Asian Earth Sciences*, 4, 4–5, 5,383-401
Tramutoli, V., Cuomo V, Filizzola C., Pergola N., Pietrapertosa, C. Assessing the potential of thermal infrared satellite surveys for monitoring seismically active areas. The case of Kocaeli (İzmit) earthquake, August 17th, 1999, *Remote Sensing of Environment*, 96, 409-426, 2005
TEPCO news releases (http://www.tepco.co.jp/en/press/corp-com/release)
Xiong P, X. H. Shen, Y. X. Bi, C. L. Kang, L. Z. Chen, F. Jing, and Y. Chen Study of outgoing longwave radiation anomalies associated with Haiti earthquake, *Nat. Hazards Earth Syst. Sci.*, 10, 2169–2178, 2010

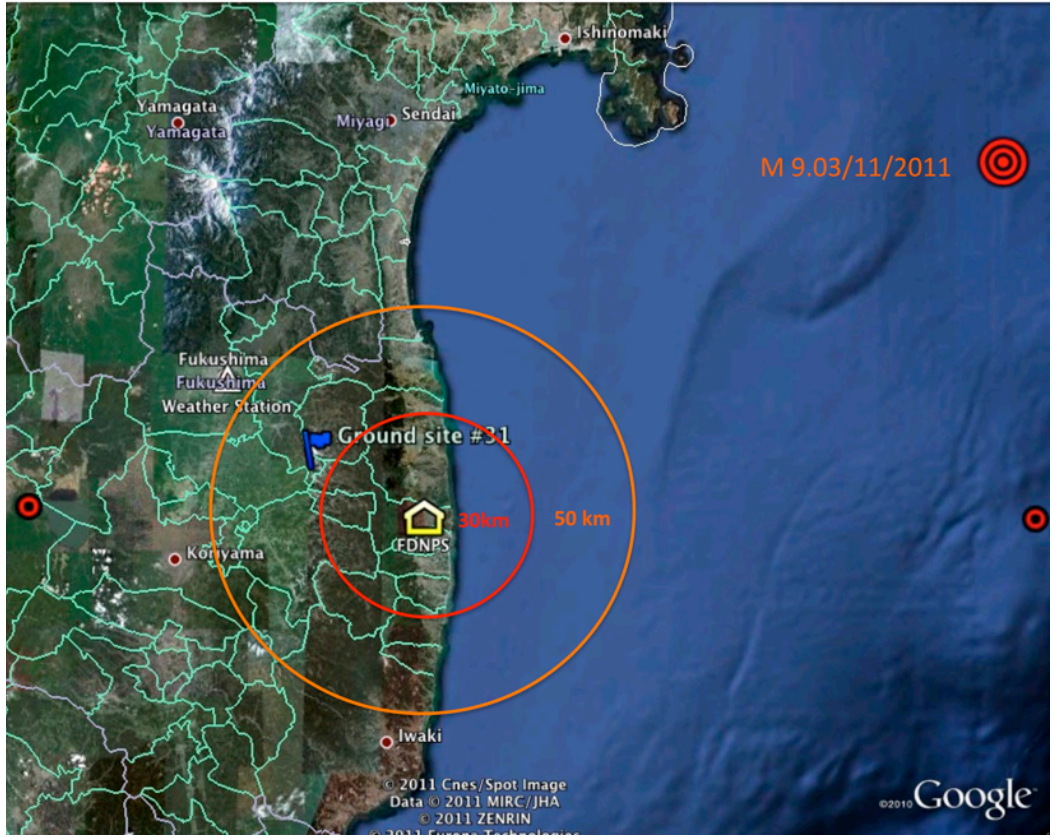

Fig.1 Google Navigation map for Southern Honshu and Fukushima-Daichi NPP (yellow house). 30km and 50km radius form FDNPP are shown with read and orange circles. M9.0 epicenter of March 11 2011 earthquake is shown with Red circles. With small red circle are shown the location of the satellite gridded points from NOAA AVHRR closes to FDNPP. With white triangle is location of the Weather stations near Fukushima city. With blue flag is the location of site#31 of ground radiometer network outside of the 30 km area.

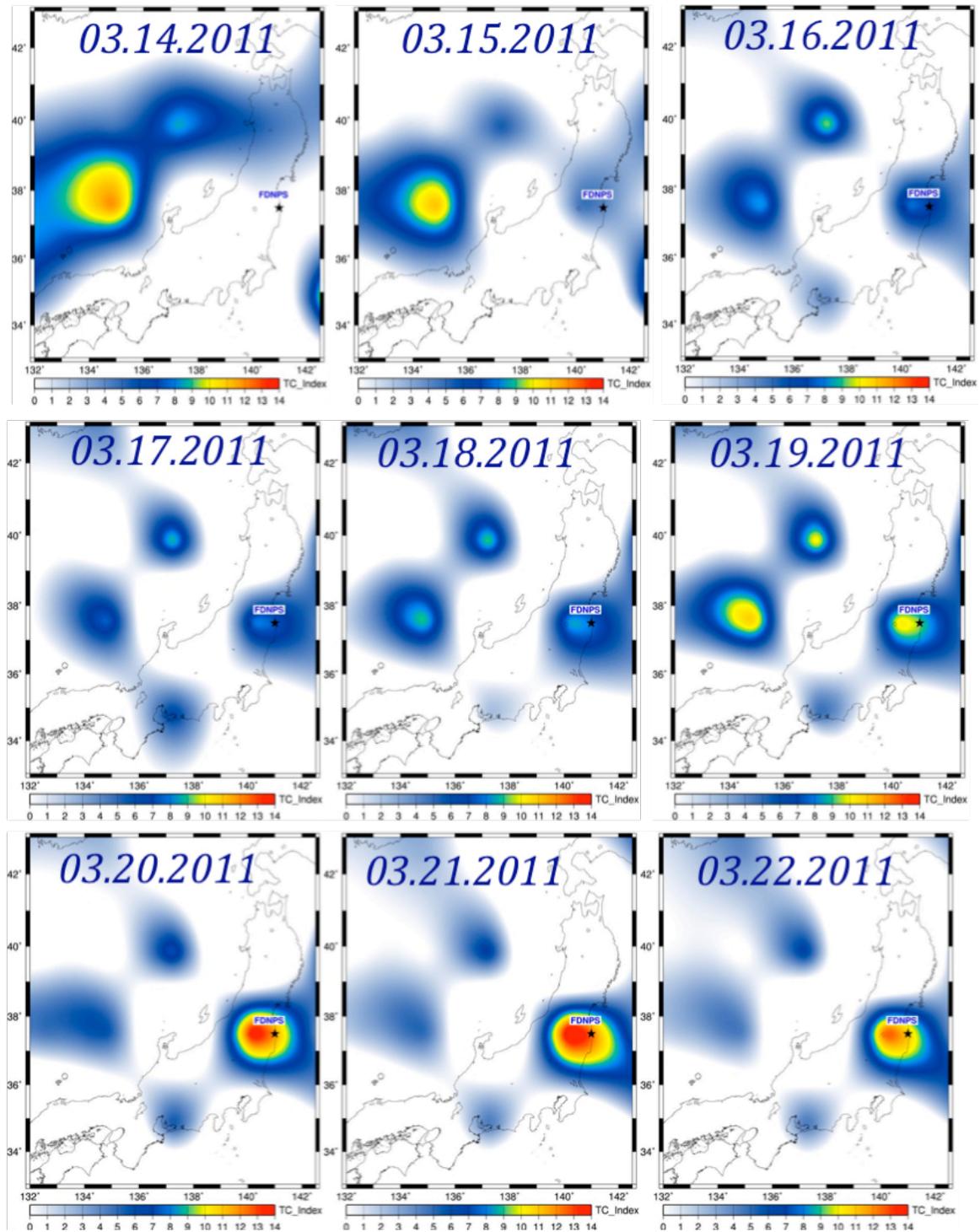
6

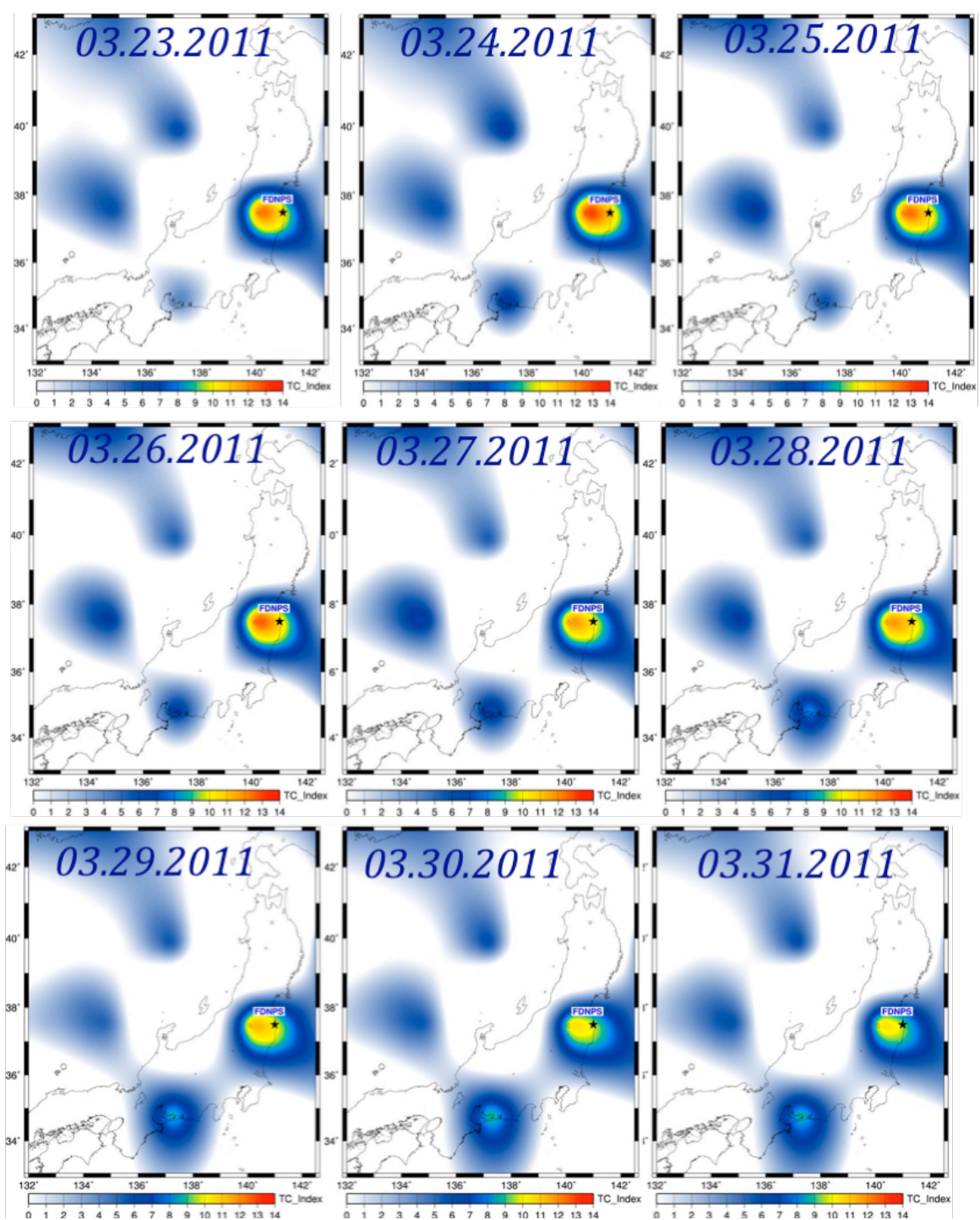

Fig.2 Time series of daytime anomalous OLR observed from NOAA/AVHRR (07.30 LT) time of the equatorial crossing) March 14-March 31, 2011 over Honshu, Japan. Tectonic plate boundaries are indicated with red lines) and major faults by brown. The location of FDNPP is indicated by a black star. The maximum rate of change for OLR been seen over FDNPP.

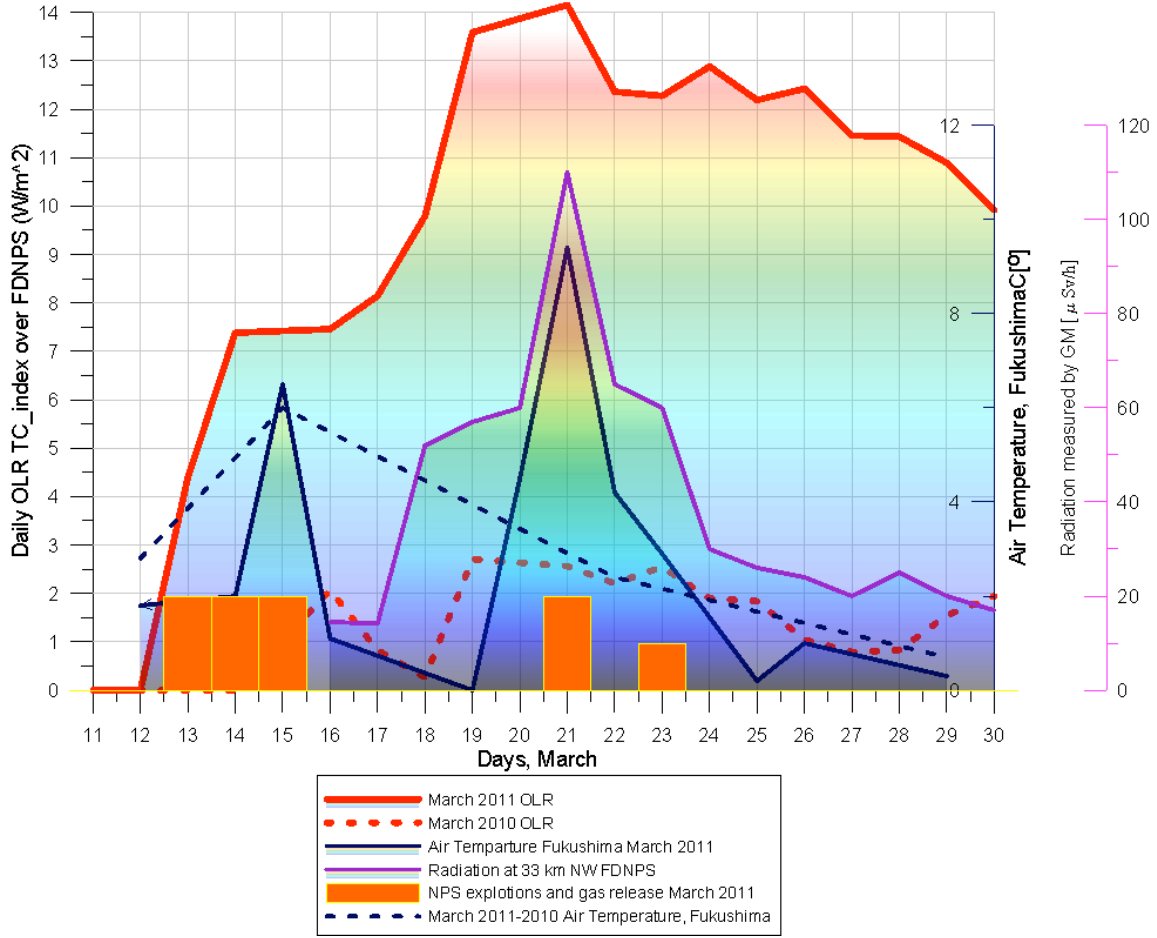

Fig.3 Satellite and ground observation during March 2011 near FDNPP: A. Accumulative graph of daily mean (07.30LT) change in OLR from NOAA/AVHRR March 11-31, 2011 (solid red).  B. Accumulative graph of daily mean (07.30LT) change in OLR from NOAA/AVHRR March 11-31, 2010 (dash red); C. March 11-30 2011 Daily Air temperature variations at 07.30LT from Fukishima weather station (solid black); D. Delta T (2011-2010) daily Air temperature variations at 07.30LT from Fukishima weather station (dash black); E. In situ radiation intensity recorded on the ground in 30km NW from FDNPP  (pink); F. NPS explosions and gas releases in atmosphere reported by TEPCO (purple bars).